\documentclass[%
aip,
amsmath,amssymb,
reprint,%
]{revtex4-2}
\usepackage{graphicx}
\usepackage{dcolumn}
\usepackage{bm}

\usepackage[utf8]{inputenc}
\usepackage[T1]{fontenc}
\usepackage{mathptmx}
\usepackage{etoolbox}

\makeatletter
\def\@email#1#2{%
	\endgroup
	\patchcmd{\titleblock@produce}
	{\frontmatter@RRAPformat}
	{\frontmatter@RRAPformat{\produce@RRAP{*#1\href{mailto:#2}{#2}}}\frontmatter@RRAPformat}
	{}{}
}%
\makeatother

\begin{document}
	
	\preprint{AIP/123-QED}
	
	\title{A Trivial Geometrical Phase of an Electron Wavefunction in a Direct Band Gap Semiconductor CdGeAs$_{2}$}
	
	\author{Vikas Saini}
	\email{vikas.saini@tifr.res.in, thamizh@tifr.res.in}
	\author{Souvik Sasmal}
	\author{Vikash Sharma}
	\author{Suman Nandi}
	\author{Gourav Dwari}
	\author{Bishal Maity}
	\author{Ruta Kulkarni}
	\author{Arumugam Thamizhavel}
	\affiliation
	{Department of Condensed Matter Physics and Materials Science, Tata Institute of Fundamental Research, Homi Bhabha Road, Colaba, Mumbai 400 005, India.}
	
	
	\date{\today}

	\begin{abstract}
		Chalcopyrite compounds are extensively explored for their exotic topological phases and associated phenomena in a variety of experiments. Here, we discuss the electrical transport properties of a direct energy gap semiconductor CdGeAs$_{2}$. The observed transverse magnetoresistance (MR) is found to be around 136$\%$ at a temperature of 1.8~K and a magnetic field of 14~T, following the semiclassical exponent MR~$\sim$~B$^{2.18}$. The MR analysis exhibits a violation of the Kohler rule, suggesting the involvement of multiple carriers in the system. Below 15~K, with decreasing magnetic field, the MR increases, leading to the well known quantum interference phenomenon weak localization (WL). The analysis of the magnetoconductivity data based on the Hikami-Larkin-Nagaoka (HLN) model unveils three dimensional nature of the WL and the weak 
		spin-orbit coupling in CdGeAs$_{2}$. The phase coherence length follows the L$_{\phi}$~$\sim$~$T^{-0.66}$ power law, which exhibits the 3D nature of the observed WL feature.                  
	\end{abstract}

	\maketitle
	In condensed matter physics the electronic phases are classified primarily based on their band gap and order of the carrier densities that lead to many phenomena and novel properties that are driven by symmetry breaking of the systems~\cite{arodz2003patterns, beekman2019introduction, mcgreevy2022generalized}. 
	On the contrary, topological phases of materials exhibit symmetry protected phases along with the combination of novel band crossings. The topological materials lead to a plethora of interesting properties like quantum Hall effect, quantum spin Hall effect, linear magnetoresistance and weak antilocalization, ultrahigh magnetoresistance, chiral anomaly, etc. which have been experimentally measured on the number of materials~\cite{liang2015ultrahigh, shekhar2015extremely, feng2015large,saini2021linear, lu2014weak}. However, quantum interference phenomena such as weak localization and weak antilocalization are also observed in the semiconducting compounds which have been explained thoroughly by the existing theories~\cite{newton2017weak, kilanski2013low}.          
	
	Chalcopyrite materials crystallize in the general chemical formula $ABC_{2}$ where $C$ represents the pnictide elements, are being studied extensively for their novel features to transport experiments. Many of the compounds of this series have been predicted and realized experimentally which appear in the whole range of the electronic phases of the band inversion to double band inversions. The ideal Weyl phase have been proposed by Ruan et al, for CuTlSe$_{2}$ and ZnPbAs$_{2}$, etc~\cite{ruan2016ideal}.  
	
	Thermoelectric properties of this series of semiconducting materials are crucial due to the sharp density of states near the Fermi level, as studied in CuGaTe$_{2}$ and Ag$_{1-x}$GaTe$_{2}$ materials~\cite{kumar2013thermoelectric, yusufu2011thermoelectric, parker2012thermoelectric}. These materials exhibit large thermopower.
	
	Here, we discuss the transport properties of a chalcopyrite material CdGeAs$_{2}$. The temperature dependence of resistivity shows a negative slope across the entire temperature range from 1.8 to 300 K, indicating the semiconducting nature of this compound. The MR is positive for temperatures above around 15~K in the magnetic field range from -14 to +14 T. At lower temperatures (less than 15~K), the MR increases as the magnetic field decreases, indicating the weak localization phenomena in CdGeAs$_{2}$. The estimated small value of $\alpha$ suggests weak spin-orbit coupling (SOC) in CdGeAs$_{2}$.   
	
	We also examined the thermopower ($S$) of CdGeAs$_{2}$ above room temperature. The negative thermopower indicates the dominant $n$-type carriers in the system, which intrinsically appear in the crystal growth process and are not due to any external doping. High values of $S$ are observed for the B1 and B2 samples at the temperatures around 530~K with the values of -185 and -285~$\mu$V/K, respectively.
	
	The material CdGeAs$_{2}$ was prepared using the modified Bridgman method by involving fast heating and natural cooling of the sample. Initially, the stoichiometric ratio of the individual elements Cd, Ge, and As was taken in an alumina crucible, which was then sealed inside a quartz tube under a vacuum of around 10$^{-5}$~mbar. The tube was placed in a box furnace at a temperature of around 500 $^\circ$C for 4 days later the furnace was switched off for natural cooling. The tiny pieces of polycrystals of CdGeAs$_{2}$ were obtained, and a small amount of crystals were isolated and crushed into powders for powder X-ray diffraction.
	
	Before performing electrical and thermal transport measurements, pellets of CdGeAs$_{2}$ were made by applying $\sim$ 5~GPa of hydrostatic pressure for nearly 5~hrs on the fine powders of CdGeAs$_{2}$ using tungsten carbide die. The pellets were then sealed under vacuum inside the quartz tube and kept at 400~$^\circ$C for 3 days for the sintering process.
	The electrical resistivity measurements were performed using the conventional four probe method, where the electrical contacts were made by the two component silver paste with gold wires attached to the sample surface. A 1.28$\times$2.68$\times$0.21 mm$^{3}$ dimensions sample was used for the resistivity measurements. The electrical transport measurements were carried out in a PPMS (Quantum Design, USA) equipped with a 14~T magnetic field, over a temperature range of 1.8 - 300~K.
	Thermopower was measured in a homemade setup where the sample was placed between the two copper plates for the thermal gradient. The experimental setup is described in reference~\cite{sk2022instrument}. 
	
	Thermopowers were calculated using the Boltzmann transport equation under the constant scattering time approximation (CSTA), as implemented in BoltzTraP2. The calculations were based on self-consistent results from the full potential linearized augmented plane wave (FP-LAPW) method implemented in WIEN2k with 10,000 $k$ points. The thermopower results were converged using 149,784 irreducible $k$ points~\cite{blaha2001wien2k,schwarz2003solid,schwarz2003dft, madsen2006boltztrap, madsen2018boltztrap2}.
	
	\begin{figure}[!]
		\includegraphics[width=0.48\textwidth]{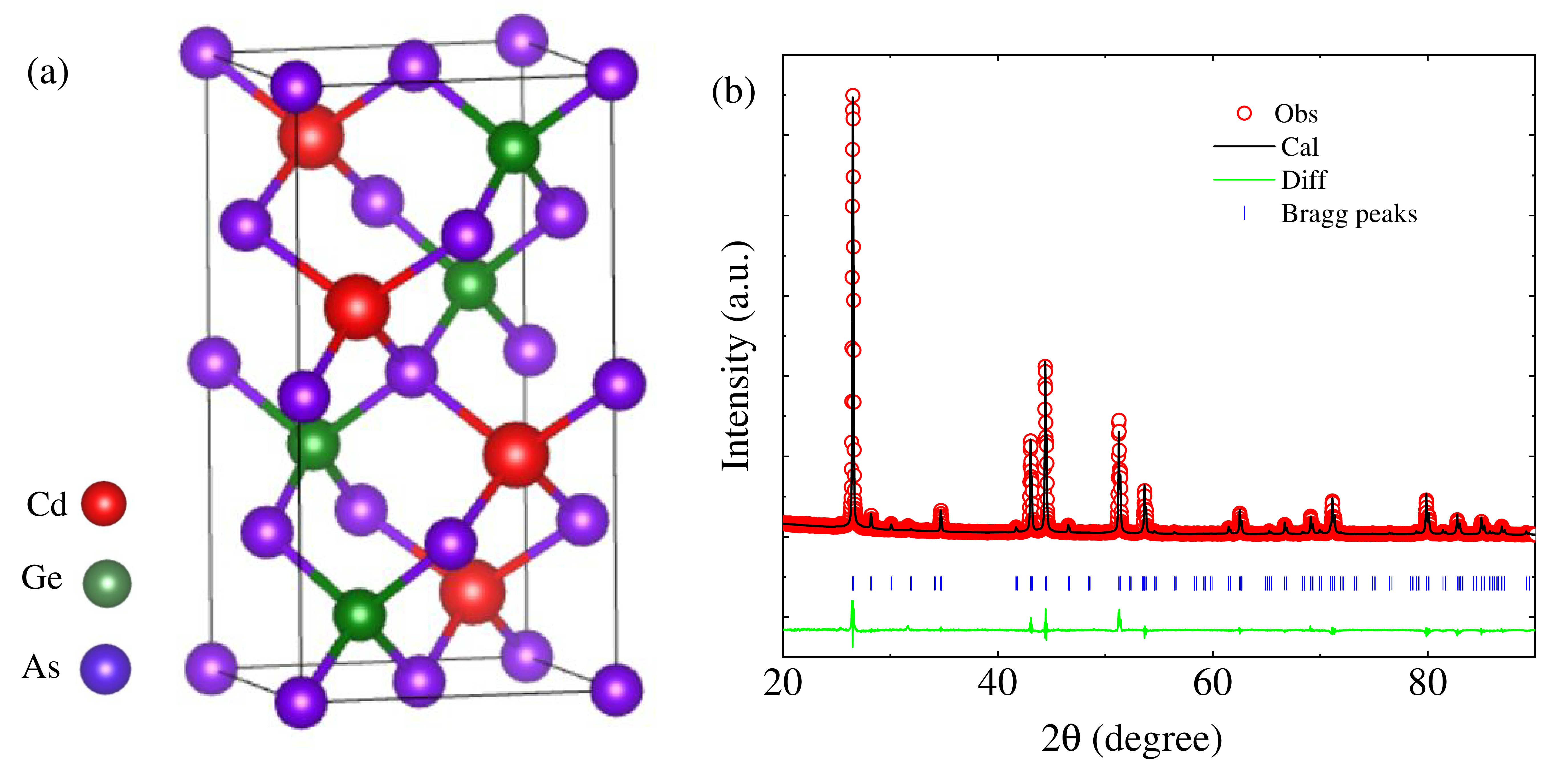}
		\caption{(a) Crystal structure of CdGeAs$_{2}$. The red, green, and purple spheres are Cd, Ge, and As atoms, respectively. (b) The powder X-ray diffraction pattern of CdGeAs$_{2}$, where red open circles correspond to the experimentally observed pattern, the black line is the refined rietveld fit, and blue ticks represent the peak positions.}
		\label{Fig1}
	\end{figure}  
	
	\begin{figure*}[!]
		\includegraphics[width=0.94\textwidth]{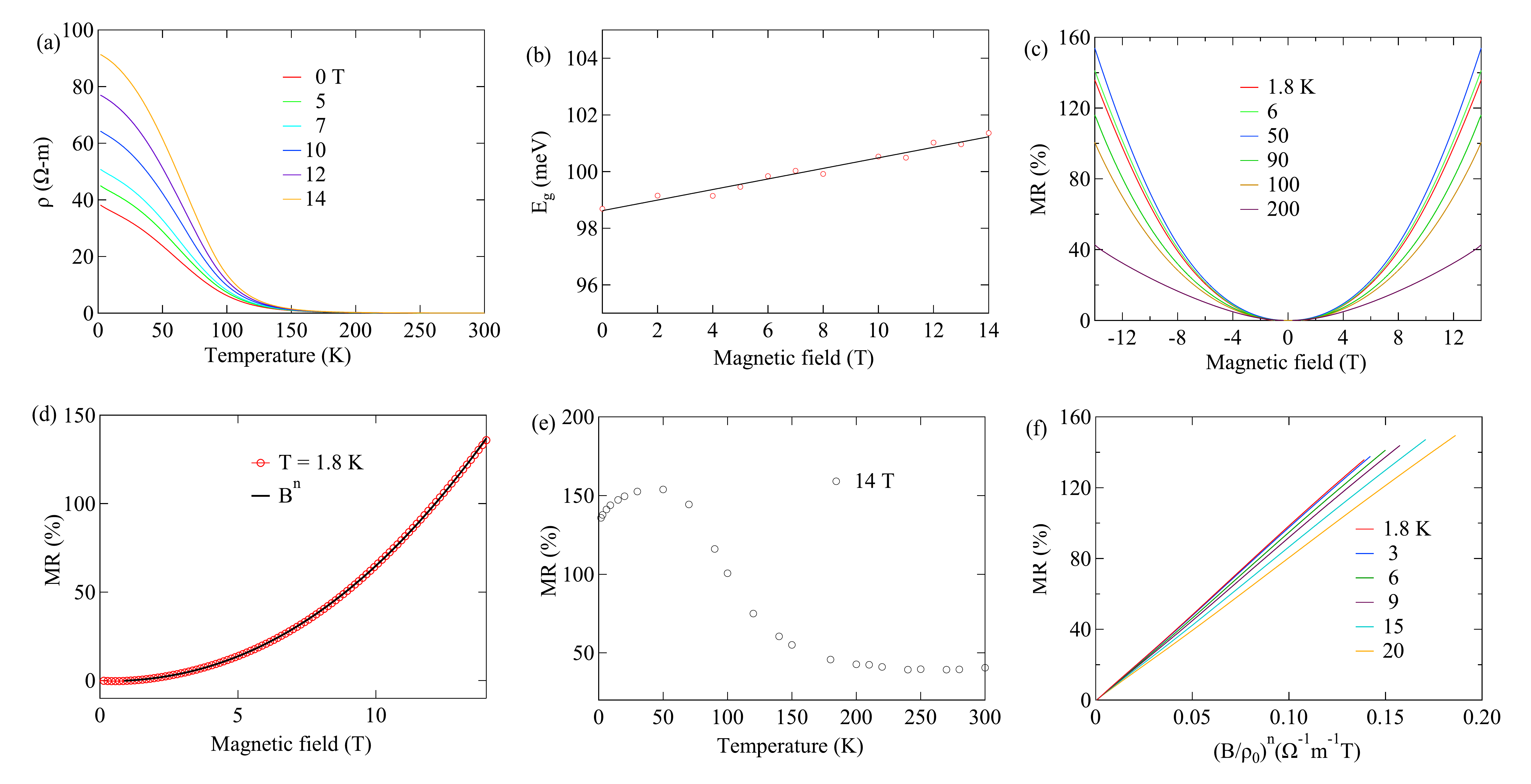}
		\caption{(a) Longitudinal resistivity as a function of temperature in applications of various static magnetic fields. (b) Thermally excited band gap as a function of the magnetic field up to 14~T. (c) Magnetoresistance (MR) versus magnetic field plot for several temperatures from 1.8 to 200 K. (d) MR plot at $T$ = 1.8~K up to 14~T, where $B^{n}$ is fitted with $n$ = 2.18. (e) MR dependence with the temperatures at 14~T magnetic field. (f) Kohler plot at various temperatures with  exponent $n$ = 2.18.}\label{Fig2}
	\end{figure*}

	\begin{figure*}[!]
		\includegraphics[width=0.94\textwidth]{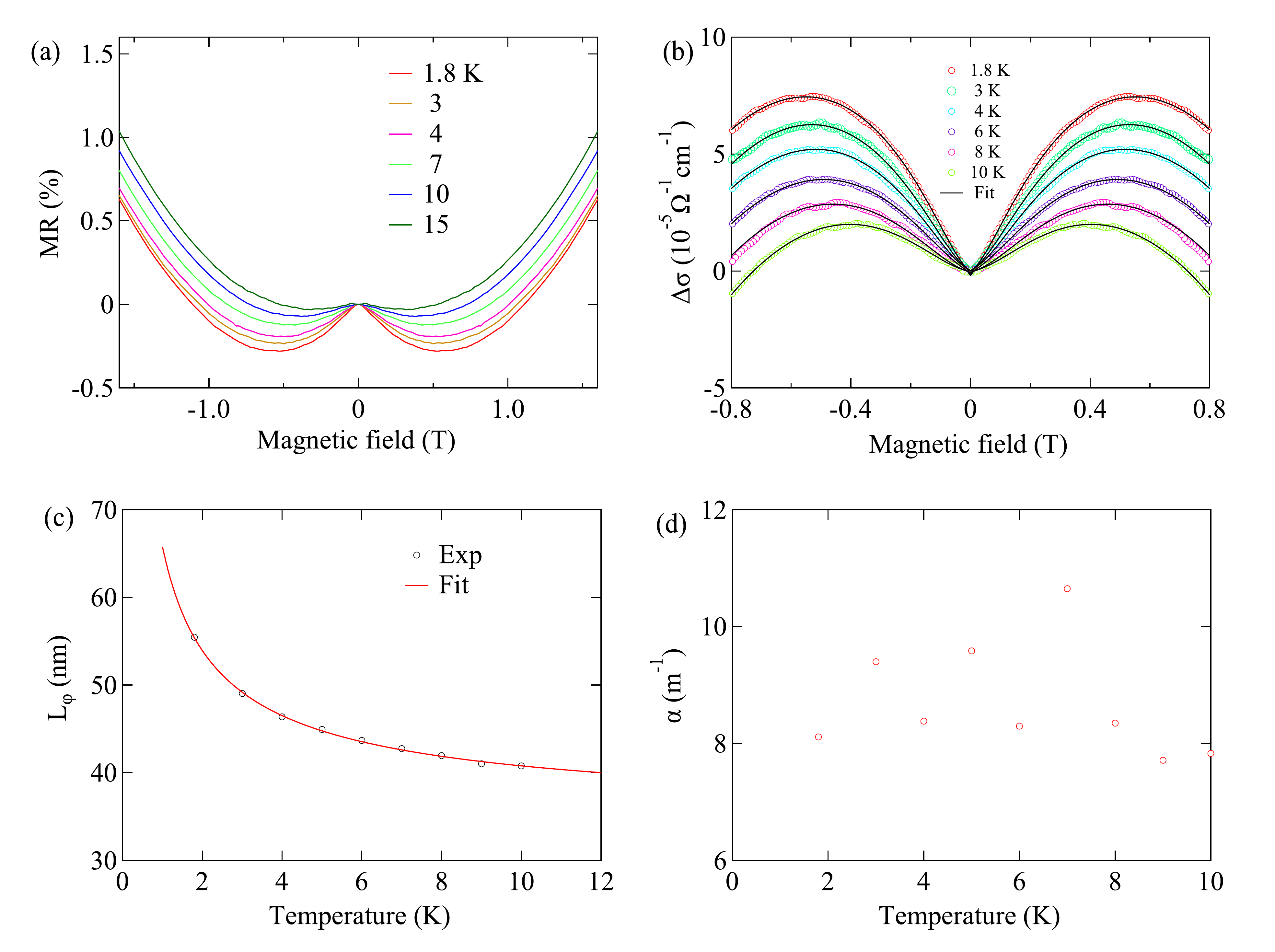}
		\caption{(a) MR plot as a function of the magnetic field, where the lower regime of magnetic field exhibits negative MR where WL is observed. (b) Modified HLN fitted normalized electrical conductivity with a magnetic field. (c) Phase coherence length ($L_{\phi}$) as a function of temperature, where the red plot is the temperature exponent fit. (d) Obtained HLN fitting parameter $\alpha$ for various temperatures.}\label{Fig3}	
	\end{figure*}

	CdGeAs$_{2}$ forms a tetragonal unit cell where CdGe layers are sandwiched between As layers along the longest crystal axis. A single unit cell consists of 9 layers. The bottom and top four layers are mirror reflections of each other in the CdGe (5$^{th}$) layer, combined with a 90$^\circ$ rotation depicts an inversion asymmetry in CdGeAs$_{2}$. However, CdGeAs$_{2}$ falls into the non-symmorphic space group category I$\bar{4}$2d (No. 122), possessing lattice translation symmetry combined with other crystalline symmetries. The atoms in a unit cell are diagonally mirror reflected in the $\pm$m$_{xy}$ planes, whereas m$_{x}$ and m$_{y}$ mirror symmetries are absent in CdGeAs$_{2}$ (see Fig.~\ref{Fig1}(a)).
	
	The powder X-ray diffraction pattern is shown in the Fig.~\ref{Fig1}(b). Experimentally observed XRD peaks match with the Rietveld refined pattern that depicts the phase purity of the prepared CdGeAs$_{2}$ sample. The refined pattern for CdGeAs$_{2}$ is depicted in the black color and the estimated lattice parameters are $a$ = $b$ = 5.941 $\rm \AA$, and $c$ = 11.214 $\rm \AA$. 
	
	We shall discuss the electrical transport results on CdGeAs$_{2}$ polycrystalline sample.    
	The resistivity $\rho$(T) of CdGeAs$_{2}$ is found to be small, nearly 0.019 $\Omega$-m at 300 K, and it starts to increase with decreasing temperature, attaining the value of 38~$\Omega$-m at 1.8~K. The slope of the resistivity in the entire temperature range is negative ($\frac{d\rho(T)}{dT} < 0$), which leads to the conclusion of the semiconducting nature of CdGeAs$_{2}$.
	In addition, the semiconducting phase of this material has been optically studied, where a direct band gap of around 0.57~eV at room temperature is observed, which further increases with lower temperatures. First principles calculations using modified Becke Johnson (mBJ) exchange potential to exhibit a direct band gap semiconductor of 0.65 eV gap, consistent with the experimental results. Thus, it supports establishing the semiconducting phase in CdGeAs$_{2}$~\cite{mccrae1997photoluminescence, akimchenko1973electroreflection, bai2005urbach}.     
	
	The electrical resistivity measured in the presence of static magnetic fields up to 14~T is depicted in Fig.~\ref{Fig2}(a). The resistivity below around 150~K starts to increase significantly with increasing magnetic field. The thermal activation gap is estimated using the Arrhenius equation $\sigma \sim \sigma_{0}$ exp(-E${g}$/k${B}$T), where $k_{\rm B}$ is the Boltzmann constant, $E_{\rm g}$ is the band gap, and $T$ is the absolute temperature.
	The thermally excited band gap is around 98.6~meV in a zero magnetic field, which is roughly 1.5 times more than the band gap value at room temperature. The application of a magnetic field slowly increases $E_{g}$ linearly with an increase in the magnetic field, reaching nearly 101.3~meV for a 14~T magnetic field, as shown in Fig.~\ref{Fig2}(b).
	
	The magnetoresistance( MR($\%$) = $\frac{\rho_{xx}(B) - \rho_{xx}(0)}{\rho_{xx}(0)}$$\times$100) of CdGeAs$_{2}$ increases with the increasing magnetic field for every temperature from around 15 to 300~K as shown in Fig.~\ref{Fig2}(c). For lower temperatures less than 15~K MR becomes negative in low field regime owing to the quantum interference phenomenon of weak localization and this feature is suppressed with increasing magnetic fields as we discuss this phenomenon later in detail. The observed value of MR at 14~T and 1.8~K is 136$\%$ and an increase in temperature also increases MR until $\sim$ 50~K that reaches up to 154$\%$. Beyond this temperature, it reduces rapidly down to 55$\%$ at 150~K and then slowly decreases until $T$ = 300~K. 
	
	According to the semiclassical Boltzmann theory, in the weak field limit ($\omega\tau << 1$), the MR follows a quadratic field dependence, and for the strong field limit, it gets saturated.
	
	However, we do not observe saturation in MR up to an applied magnetic field of 14~T. Therefore, we can use a semiclassical description of the unsaturated MR by employing MR $\sim$ B$^n$, where $n$ = 2.18 which is shown in the Fig.~\ref{Fig2}(c).
	
	We observed that the MR plots for various temperatures do not overlap onto each other when plotted against ($\frac{B}{\rho_{0}}$)$^{n}$, which implies the violation of the Kohler rule. This suggests the presence of multiple carriers in CdGeAs$_{2}$.
	The Kohler plot is depicted in the Fig.~\ref{Fig2}(f) in the temperature range 1.8 - 20~K, where $n$ = 2.18 is the obtained exponent of MR.   
	
	The low magnetic field data up to a temperature of 10~K is shown in Fig.~\ref{Fig3}(a), which reveals negative MR down to 1.12$\%$ for $T$ = 1.8 K. At lower magnetic fields, the MR increases, suggesting the weak localization (WL) in CdGeAs$_{2}$. However, this effect disappears at temperatures above 15~K and the MR becomes positive in the entire range of the magnetic field.
	
	We observed WL phenomenon in the CdGeAs$_{2}$. As we mentioned earlier that from the semiclassical theory MR increases quadratically in the weak limit and it gets saturated in the strong fields. But the quantum corrections in the low field regime leads to the quantum interference phenomena as WL and WAL depending on the interference between the time reversal invariant paths of electron wavefunctions as widely observed for the number of crystalline systems. For this quantum phenomenon, the phase coherence length ($L_{\phi}$) must be greater than the mean free path ($l$) that results in the interference between time reversal (TR) paths of the electron wavefunctions that essentially change the electrical conductivity of the measured compounds. If the phase difference between the TR paths is constructive then it results in more backscatterings in the system thus leads to the suppression of electrical conductivity from the normal classical behavior that is known as the WL. On the other hand, destructive interference gives a $\pi$ phase between the TR paths that suppresses the backscatterings in the system and hence leads to an increase in electrical conductivity that is defined as  WAL. The increase in temperature causes a decrease in $L_{\phi}$ which essentially suppresses these features. This whole phenomenon appears in the time reversal invariant condition therefore, it is observed in the lower magnetic field regime and the application of the high magnetic field suppresses this feature.    
	
	To observe WAL, the strong spin-orbit coupling is required to induce a nontrivial phase that increases electrical conductivity. This effect can be two- or three-dimensional in nature.
	The Hikami-Larkin-Nagaoka (HLN) model is commonly used to explain WL and WAL in two dimensional materials and has been applied to identify surface states in topological insulators~\cite{lu2014weak, Liu}.                   
	
	To understand the weak localization in CdGeAs$_{2}$ we employed the HLN model to analyze the experimental data that alone does not produce the good fittings with the experimental data of normalized magnetoconductivity ($\Delta$$\sigma$ = $1/\rho$(B) - $1/\rho$(0)); where $\rho$ is the electrical resistivity. However, the fit was not good. This suggested classical contributions of the cyclotron orbits should also be taken into account. Hence we added the additional $B^{2}$ term to obtain a reasonably good fit of the magnetoconductivity data.
	Further, we have incorporated the three dimensional term of the weak localization $B^{1/2}$ which also improved the HLN fitting. 
	Our estimated $\alpha$ parameter from the modified HLN model is around 8 m$^{-1}$ at 1.8~K which is reasonably lower than the observed values in the three dimensional nature of LuPdBi and LuPtSb~\cite{xu2014weak, hou2015transition}. However, the lower value of $\alpha$ could be attributed to the semiconducting nature of the material and observed WL suggests the low spin-orbit coupling strength in CdGeAs$_{2}$ as observed in the annealed semiconducting material PbPdO$_{2}$~\cite{choo2015crossover}.  
	
	The fitted normalized magnetoconductivity $\Delta \sigma$ is depicted in the Fig.~\ref{Fig3}(b) for various temperatures up to 10~K. The modified HLN equation is expressed as follows~\cite{assaf2013linear,wang2020weak}. 
	\begin{equation}
		\label{Eq1}
		\Delta \sigma (B) = \frac{\alpha e^{2}}{2 \pi^{2} \hbar} [\Psi(\frac{1}{2}+\frac{\hbar }{4 e B L_{\phi}^{2}})-ln(\frac{\hbar }{4 e B L_{\phi}^{2}})]-\beta B^{2} + \delta B^{1/2}
	\end{equation}             
	
	The obtained phase coherence length $L_{\phi}$ and $\alpha$ parameters are shown in the Fig.~\ref{Fig3}(c-d). 
	The phase coherence length relation with temperature can be expressed as follows:    
	$\frac{1}{L^2_{\phi}(T)}$ = $\frac{1}{L^2_{\phi}(0)}$ + A$_{ee}$T$^{p'}$ + A$_{ep}$T$^{p''}$, where constants A$_{ee}$ and $A_{ep}$ reflects the electron electron and electron phonon interactions. 
	For the electron-electron interaction dominating system exponent $p'$ possesses 1 and 3/2 values for the two and three-dimensional nature of WL, respectively. Since in CdGeAs$_{2}$ the electron-electron interactions are dominant compared to the phonon-phonon interactions, the above expression can be simplified as L$_{\phi}$ $\sim$ T$^{-p}$ where $p = p'/2$. 
	The temperature dependence of the $L_{\phi}$ results in the exponent $p$ = 0.66 (Fig.~\ref{Fig3}(c)), which is close to the expected value of $p = p'/2= 0.75$ suggesting the three-dimensional nature of the weak localization.
	
\begin{figure}[!]
		\includegraphics[width=0.48\textwidth]{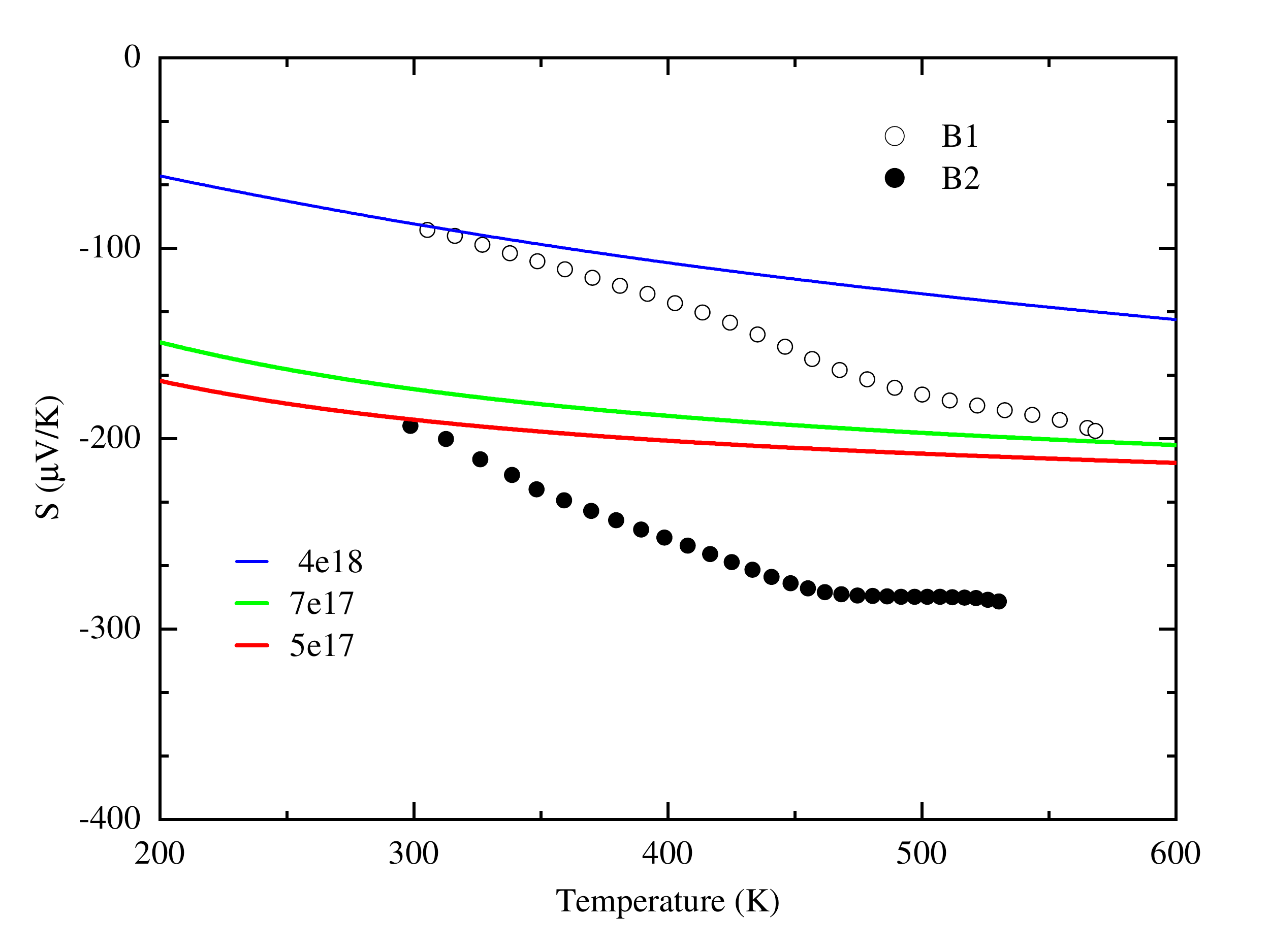}
		\caption{Thermopower ($S$) is plotted against the temperature from 200 - 600~K. Red, green, and blue curves are the calculated $S$ for the different carrier densities. Open and closed black circles are the experimentally observed $S$ for two samples.}
		\label{Fig4}
	\end{figure}

	The thermopower ($S$) of CdGeAs$_{2}$ was measured on two samples B1 and B2 from temperature $\sim$300 to 570~K. The magnitude of $S$ increases with temperature, as shown in Fig.\ref{Fig4}, and the negative values indicate natural $n$-type doping in CdGeAs$_{2}$. The qualitative behavior of both samples is similar, with a subtle shift in values that may be due to differences in carrier densities. As shown in Fig.~\ref{Fig4}, the calculated $S$ for three different carrier densities reveals that increasing carrier density leads to a decrease in $|S|$, thus the carrier density of sample B2 should be smaller than that of sample B1. The magnitude of $S$ observed falls within the range predicted for $n$-type dopings.
	The thermopower calculations are performed under the constant relaxation time approximation (CSTA)~\cite{madsen2018boltztrap2}.
	The WL phenomena is generally related to the impurities/disorders present in the sample. Here in this case of CdGeAs$_{2}$ the observed WL is attributed to the disorders/impurities. Further, it has to be pointed out that the possible scatterings from the impurities near the Fermi level are not considered in calculations therefore the difference between observed and calculated thermopower is expected for CdGeAs$_{2}$.
	
	The effects of impurities and imperfections in crystal on observed thermopower can be understood with the Mott relation~\cite{he2017advances} given in the  Eq.~\ref{Eq2} 
	
	\begin{equation}
		S(n,T) =  \frac{\pi^{2} k_{B}^{2}T}{3q} \left[\frac{1}{n}\frac{dn(E)}{dE} + \frac{1}{\mu}\frac{d\mu(E)}{dE}\right]_{E_{F}} 
		\label{Eq2}
	\end{equation} 
	
	Here $k_{B}$ is the Boltzmann constant and $q$ denotes the charge of carriers, n(E) and $\mu$(E) represents the carrier density and mobility of the charge carriers. $T$ stands for the absolute temperature. 
	The first term in the Eq.~\ref{Eq2} represents the contributions from the density of states at the Fermi level, the sharp increase in density of states against energy leads to the large thermopower whereas the second term appears from the scatterings caused by disorders that have not been attributed in the calculated thermopower. It is likely that the nonzero second term in Mott relation may attribute to the difference between the calculated and experimentally measured thermopowers on CdGeAs$_{2}$.     
	
	In conclusion, we studied the transport properties of a semiconducting chalcopyrite system CdGeAs$_{2}$ which exhibits weak localization phenomenon.  
	For temperatures below $\sim$150~K, semiclassical dependence of the magnetoresistance is observed.
	The phase coherence length ($L_{\phi}$) exponent with temperatures results in the three-dimensional nature of the WL in CdGeAs$_{2}$. Additionally, a large thermopower is measured for CdGeAs$_{2}$ as a consequence of the sharp increase in the density of states.  
	
	The authors gratefully acknowledge high temperature thermoelectric facility at IIT Mandi, India for the thermopower measurements. VS and AT would like to specially thank Prof. Sudhir Kumar Pandey and MR. Shamim Sk for their assistance in the thermopower measurements.
	
	\section*{Data Availability Statement}
	The data that support the findings of this study are available
	from the corresponding author upon reasonable request.
	

\section*{References}
	\nocite{*}
	\providecommand{\noopsort}[1]{}\providecommand{\singleletter}[1]{#1}%
	%


\end{document}